# Alignment of Cybersecurity Incident Prioritisation with Incident Response Management Maturity Capabilities

Abdulaziz Gulay, Leandros Maglaras

Edinburgh Napier University, 10 Colinton Road., Edinburgh, EH10 5DT, UK

**Abstract**

The increasing frequency and sophistication of cybersecurity incidents pose significant challenges to organisations, highlighting the critical need for robust incident response capabilities. This paper explores a possible utilisation of IR CMMs assessments to systematically prioritise incidents based on their impact, severity, and the incident response capabilities of an organisation in specific areas associated with human and organisational factors. The findings reveal common weaknesses in incident response, such as inadequate training and poor communication, and highlight best practices, including regular training programs, clear communication protocols, and well-documented response procedures. The analysis also emphasises the importance of organisational culture in enhancing incident response capabilities. By addressing the gap in understanding how the output of IRM assessments can be immediately utilised to prioritise high-risk incidents, this paper contributes valuable insights to academia and practice, offering a structured approach to enhancing organisational resilience against cybersecurity threats.

1. Introduction

The frequency and sophistication of cybersecurity incidents have escalated over the past decade, posing significant challenges to organisations worldwide. Cyber threats are evolving rapidly, becoming more complex and challenging to detect. Organisations of all sizes and across various industries are increasingly vulnerable to these threats, which can cause substantial financial losses, reputational damage, and operational disruptions. High-profile breaches, such as those at Target Corp. in 2013, Sony Pictures Entertainment in 2014, and Equifax in 2017, have highlighted the critical importance of robust incident response capabilities. These incidents exposed the vulnerabilities in the cybersecurity defences of these organisations but also highlighted the inadequacies in their incident response strategies.

Despite the increased emphasis on cybersecurity, many organisations struggle to manage and respond to cyber incidents effectively. The breaches at Sony Pictures Entertainment, Target Corp., and Equifax underscored significant shortcomings in incident response, particularly around human and organisational factors. These factors include the adequacy of training and awareness programs, the effectiveness of communication and coordination during incidents, and the organisation's overall preparedness to handle cyber threats.

While various Incident Response (IR) Capability Maturity Models (CMMs) have been developed to assess and enhance organisational readiness, there remains a gap in understanding how the outcomes of these assessments can be used to identify and prioritise high-risk incidents immediately. This research aims to address this gap by exploring the use of IR capability maturity assessments, focusing specifically on the human and organisational factors that influenced the response to the breaches mentioned above.

The primary objective of this paper is to investigate how IR capability maturity assessments can be leveraged to identify high-priority incidents effectively. To achieve this outcome, the study will examine the existing maturity models used for incident response assessment and explore their application in incident prioritisation. The goal is to provide actionable insights and practical recommendations for organisations to enhance their incident response capabilities and better manage cybersecurity risks.

Examination of existing research done on IR CMMs

Cyber incident handling failures represent complex challenges influenced by various factors such as concealment of incidents, inadequate training and awareness, information security governance, incident handling best practices, risk management, lack of coordination and communication, and organisational culture in federated environments. These factors interact in complex ways, making organisations need to adopt a comprehensive approach to incident response.

A crucial step towards addressing these challenges in any organisation is the assessment of their IR capability maturity through a Capability Maturity Model (CMM). Several published MMs target IT Security; however, only a few are focused and dedicated to IRM (Bitzer et al., 2023).

The main objective is to draw on the findings of (Bitzer et al., 2023) in exploring IR CMMs that integrate organisational processes and human factors, reflecting on the areas of failure mentioned in the previous sections and the interconnected nature of cybersecurity.

(Bitzer et al., 2023) presented a comprehensive evaluation of a few IR CMMs, where several shortcomings were revealed in the literature review that rendered some models to be considered either not applicable to all organisations, not available for free, complicated, or lacking the socio-technical capability element.

There appears to be a notable gap in the literature explicitly linking incident prioritisation directly with the outcomes of IR capability maturity assessments. While extensive research exists on incident response, maturity models, and prioritisation techniques separately, integrating these concepts into a cohesive framework is less commonly explored.

Maturity models such as NIST, CMMI, IRM3, and CERT-RMM provide guidelines for assessing the maturity of incident response management. However, there is limited research on systematically using the results of these assessments to prioritise incidents.

Detailed methodologies that convert maturity assessment outcomes into quantitative metrics influencing incident prioritisation are sparse. Research could focus on developing algorithms or scoring systems linking maturity levels with prioritisation criteria.

More empirical and case studies are needed to demonstrate how organisations successfully integrate maturity model assessments with incident taxonomy frameworks. Real-world examples could provide practical insights and validate theoretical models.

Current literature often treats IR capability maturity and prioritisation as separate components. A more holistic approach is needed to combine these elements into a unified incident management strategy, potentially through comprehensive guidelines that align maturity assessment outcomes with prioritisation protocols.

The development of dynamic models adapting to changes in an organisation's maturity level and threat landscape over time is another area with potential for further research. Such models would help organisations refine their incident prioritisation processes as their incident response capabilities evolve.

## 2. Theoretical Framework

According to (Whitman & Mattord, 2011), technology alone is not a solution to cybersecurity issues; without proper processes and trained personnel to manage and use this technology, its effectiveness is significantly limited. Including human and organisational factors in a comprehensive cybersecurity strategy is fundamental, ensuring that technology is effectively supported by robust processes and well-trained people (Whitman & Mattord, 2011).

### 2.1 Theoretical Foundations of IR Capability Maturity Assessments

IRM assessments are critical for organisations to evaluate and enhance their capability to handle cybersecurity incidents effectively. Several theoretical concepts underpin IRM assessments, including capability maturity and risk management. These concepts provide a framework for understanding how organisations can effectively manage and respond to cybersecurity incidents.

Capability maturity models (CMMs) are frameworks used to assess and improve the maturity of an organisation's processes. These models provide a structured approach to process improvement, helping organisations move from ad-hoc, reactive processes to optimised, proactive ones. The CMMs for incident response outline several maturity levels, from initial and managed stages to defined, quantitatively managed and optimising stages. Higher maturity levels indicate more sophisticated and effective incident response capabilities, where processes are well-documented, measured, and continuously improved (Caralli et al., 2011).

Although CMMs for incident response cover several aspects related to incident response, the following are the seven areas of focus in this study based on socio-organisational failure attributes identified:

- Risk Management: Processes for identifying, assessing, and mitigating risks.
- Incident Handling Best Practices: Standard operating procedures and best practices for incident response.
- Training and Awareness: Programs and initiatives to educate staff on incident response.
- Adequate Staffing: Availability of skilled personnel to handle incidents.
- Information Security Governance: Policies, procedures, and governance structures in place.
- Internal and External Communication: Effectiveness of communication channels during incidents.
- Information Security Culture: Organisational culture regarding information security practices.

The theoretical underpinnings of the following CMMs reveal their potential to address the human and organisational factors.

**Cybersecurity Capability Maturity Model (C2M2)**

- Risk Management: C2M2 emphasises structured risk management processes, which include risk assessment, mitigation, and continuous monitoring. These processes help organisations proactively identify and address potential threats (DOE, 2014).
- Cyber Security Incident Handling Best Practices: The model incorporates incident management as one of its core domains, providing best practices for incident detection, response, and recovery. This ensures that organisations are prepared to handle incidents effectively (DOE, 2014).
- Training and Awareness: C2M2 promotes the development of a cybersecurity-aware culture through regular training and awareness programs. This ensures that all personnel understand their roles in maintaining security and responding to incidents (DOE, 2014).
- Inadequate Staffing: The model focuses on building capacity within the organisation by identifying skill gaps and ensuring sufficient and appropriately skilled personnel are available to manage cybersecurity incidents (DOE, 2014).
- Information Security Governance: C2M2 integrates governance practices to align security policies with business objectives, ensuring that cybersecurity efforts support the organisation's overall goals (DOE, 2014).
- Internal and External Communication: The model emphasises the importance of clear communication channels for incident reporting and response. This ensures that information flows effectively within the organisation and with external stakeholders (DOE, 2014).
- Information Security Culture: C2M2 encourages a culture of security awareness throughout the organisation, promoting best practices and continuous improvement in cybersecurity (DOE, 2014).

**Cybersecurity Maturity Model Certification (CMMC)**

- Risk Management: CMMC requires organisations seeking to do business with the U.S. Department of Defense to implement risk management practices tailored to their maturity level. These practices range from basic to advanced risk management strategies, ensuring that organisations can manage risks effectively at each stage of maturity (DoD, 2020).
- Cyber Security Incident Handling Best Practices: The model specifies comprehensive best practices for incident response, including preparation, detection, analysis, containment, eradication, and recovery. This ensures that organisations follow best practices for handling cybersecurity incidents (DoD, 2020).
- Training and Awareness: CMMC mandates regular training and awareness programs tailored to the specific maturity level of the organisation. This ensures that personnel are prepared for incident handling and understand their roles and responsibilities (DoD, 2020).
- Inadequate Staffing: The model includes specific requirements for adequate staffing, ensuring that organisations have the personnel to manage cybersecurity risks and respond to incidents effectively (DoD, 2020).
- Information Security Governance: CMMC integrates governance principles to align security practices with organisational goals and regulatory standards. This ensures that cybersecurity efforts are well-managed and compliant with requirements (DoD, 2020).
- Internal and External Communication: The model ensures that communication protocols are in place to effectively manage incident reporting within the organisation and with external parties (DoD, 2020).
- Information Security Culture: CMMC promotes a compliance-driven security culture, encouraging organisations to adhere to best practices and regulatory requirements (DoD, 2020).

**NIST Cybersecurity Framework (CSF)**

- Risk Management: NIST CSF provides detailed guidelines for identifying, assessing, and managing cybersecurity risks. These guidelines help organisations align their risk management efforts with business objectives (NIST, 2018).
- Cyber Security Incident Handling Best Practices: The framework emphasises the Respond and Recover functions, outlining best practices for incident response, including communication, mitigation, and improvement. This ensures effective handling of cybersecurity incidents (NIST, 2018).
- Training and Awareness: NIST CSF recommends continuous training and awareness programs to ensure that all personnel know their roles in incident response. This promotes a well-prepared workforce (NIST, 2018).
- Inadequate Staffing: While the framework addresses resource allocation, it is less focused on the specifics of staffing. However, it does emphasise the need for adequate staffing to manage cybersecurity risks effectively (NIST, 2018).
- Information Security Governance: The framework aligns cybersecurity efforts with business and regulatory requirements, promoting strong governance practices to manage cybersecurity effectively (NIST, 2018).
- Internal and External Communication: NIST CSF strongly focuses on effective communication during incidents, ensuring that all relevant stakeholders are informed and involved in the response process (NIST, 2018).
- Information Security Culture: The framework integrates security into organisational decision-making processes, promoting a culture prioritising cybersecurity (NIST, 2018).

**CERT Resilience Management Model (CERT-RMM)**

- Risk Management: CERT-RMM integrates comprehensive risk management practices within the broader context of operational resilience. This includes risk assessment and mitigation strategies that enhance the organisation's ability to handle incidents (Caralli et al., 2011).
- Cyber Security Incident Handling Best Practices: The model focuses on best practices for incident management and recovery, ensuring that critical operations are maintained during disruptions (Caralli et al., 2011).

- Training and Awareness: CERT-RMM emphasises continuous learning and a resilience culture. It integrates training and awareness programs to ensure all personnel are prepared for incident handling (Caralli et al., 2011).
- Inadequate Staffing: The model strongly emphasises resource management and capability development, ensuring sufficient and skilled personnel are available to manage incidents (Caralli et al., 2011).
- Information Security Governance: CERT-RMM integrates governance principles for comprehensive security management, ensuring that cybersecurity efforts are aligned with organisational goals (Caralli et al., 2011).
- Internal and External Communication: The model promotes effective communication strategies during incidents, ensuring that information is shared appropriately within the organisation and with external stakeholders (Caralli et al., 2011).
- Information Security Culture: CERT-RMM emphasises the development of a resilient culture that supports continuous improvement and effective incident response (Caralli et al., 2011).

**ISO/IEC 27035**

- Risk Management: ISO/IEC 27035 provides a structured approach to managing information security risks, including those related to incidents. This ensures that organisations can effectively identify, assess, and mitigate risks (Line, 2016).
- Cyber Security Incident Handling Best Practices: The standard details a phased approach to incident management, including planning, detection, response, and lessons learned. This ensures that best practices are followed throughout the incident lifecycle (Line, 2016).
- Training and Awareness: ISO/IEC 27035 mandates continuous training and awareness programs to ensure that all personnel are equipped to handle incidents effectively. This promotes a well-prepared workforce (Line, 2016).
- Inadequate Staffing: The standard ensures that organisations have the necessary human and technical resources to manage incidents, addressing the issue of inadequate staffing (Line, 2016).
- Information Security Governance: ISO/IEC 27035 aligns incident management with organisational goals and regulatory requirements, promoting strong governance practices (Line, 2016).
- Internal and External Communication: The standard provides structured communication protocols to ensure adequate information flow during incidents. This includes internal and external communication strategies (Line, 2016).
- Information Security Culture: ISO/IEC 27035 promotes a security-conscious culture within the organisation, encouraging continuous improvement and adherence to best practices (Line, 2016).

**ENISA's Incident Management Maturity Model (IM3)**

- Risk Management: ENISA's IM3 focuses on developing robust risk management practices tailored to the needs of CSIRTs. This includes risk identification, assessment, and mitigation strategies to enhance incident response capabilities (ENISA, 2022).
- Cyber Security Incident Handling Best Practices: The model is based on the SIM3 framework and provides detailed guidelines for incident handling across three maturity tiers. This ensures that best practices are followed for incident management (ENISA, 2022).
- Training and Awareness: IM3 emphasises the importance of training and awareness programs for CSIRT members. This ensures that all personnel are prepared for incident handling and understand their roles and responsibilities (ENISA, 2022).

- Inadequate Staffing: While IM3 focuses on preparedness, it places less emphasis on the specifics of staffing. However, it ensures that CSIRTs are adequately prepared to manage incidents (ENISA, 2022).
- Information Security Governance: The model includes principles that ensure incident response activities align with organisational and regulatory requirements. This promotes strong governance practices (ENISA, 2022).
- Internal and External Communication: IM3 guides effective communication during incidents, ensuring that information is shared appropriately within the CSIRT and with external stakeholders (ENISA, 2022).
- Information Security Culture: The model emphasises building a security-aware culture within CSIRTs, promoting best practices and continuous improvement in incident management (ENISA, 2022).

## 2.2 Comparison of CMMs

Furthermore, considering the theoretical underpinnings of the CMMs in the previous sub-section, the following table shows a comparison of the effectiveness of each CMM in assessing the seven specific areas of interest identified in the literature review.

The scoring is on a scale of one to five, where one marks the least effective and five shows the most effective.

| Area of Capability Model | Risk Management | Cyber Security Incident Handling Best Practices | Training and Awareness Around Incident Handling | Adequate Staffing | Information Security Governance | Internal and External Communication | Information Security Culture within the Organisation |
|---|---|---|---|---|---|---|---|
| CERT Resilience Management Model (CERT-RMM) | 5 | 5 | 4 | 4 | 5 | 4 | 4 |
| ISO/IEC 27035 | 4 | 5 | 3 | 3 | 4 | 4 | 3 |
| NIST Cybersecurity Framework (CSF) | 4 | 4 | 3 | 3 | 5 | 3 | 4 |
| Cybersecurity Capability Maturity Model (C2M2) | 5 | 4 | 4 | 4 | 4 | 3 | 4 |
| Cybersecurity Maturity Model Certification (CMMC) | 4 | 4 | 4 | 3 | 5 | 3 | 4 |
| ENISA's Incident Management Maturity Model (IM3) | 3 | 4 | 5 | 4 | 4 | 5 | 4 |

*Table 1: Comparison of CMMs against categories of failures*

Below is an explanation of how each CMM performs per area of focus, reflecting each's strengths and weaknesses.

**A. Risk Management**

- **CERT-RMM** and **C2M2** score highly (5/5) because they offer comprehensive frameworks integrating risk management with operational resilience and capability building. CERT-RMM focuses on managing operational resilience by linking risk management to business continuity and incident response, which is critical in mitigating risks effectively (Caralli et al., 2011). C2M2 emphasises continuous risk assessment and improvement, helping organisations develop robust risk management practices (DOE, 2014).
- **NIST CSF** scores slightly lower (4/5) as it provides a solid framework for risk management, emphasising the identification and management of cybersecurity risks but requiring further adaptation to specific organisational needs (NIST, 2018).
- **CMMC** also scores well (4/5), particularly in regulated environments, by enforcing compliance-driven risk management practices that ensure adherence to best practices and standards (DoD, 2020).
- **ISO/IEC 27035** and **ENISA IM3** score lower (3/5) because they primarily focus on incident management rather than holistic risk management. They offer guidance on risk assessment as part of incident management but are less comprehensive in this area compared to the other models (Line, 2016; ENISA, 2022).

B. Incident Handling Best Practices:

- **CERT-RMM** and **ISO/IEC 27035** achieve top scores (5/5) as they provide detailed, structured approaches to incident handling, ensuring comprehensive coverage from planning to recovery (Caralli et al., 2011; Line, 2016).
- **NIST CSF**, **C2M2**, **CMMC**, and **ENISA IM3** score slightly lower (4/5). While these models offer robust incident handling practices, they focus on broader organisational capabilities or require further elaboration to fit incident-specific scenarios (NIST, 2018; DOE, 2014; DoD, 2020; ENISA, 2022).

C. Training and Awareness:

- **ENISA IM3** scores the highest (5/5) due to its strong emphasis on continuous training and awareness, particularly within CSIRTs. The model encourages regular drills, simulations, and knowledge sharing to maintain a high level of readiness (ENISA, 2022).
- **CERT-RMM**, **C2M2**, and **CMMC** score well (4/5) because they also emphasise the importance of training and awareness but are slightly less focused on continuous training than ENISA IM3. These models integrate training within broader organisational resilience and maturity frameworks (Caralli et al., 2011; DOE, 2014; DoD, 2020).
- **ISO/IEC 27035** and **NIST CSF** score lower (3/5) as their guidance on training and awareness is less detailed, focusing more on incident handling and governance rather than ongoing staff development (Line, 2016; NIST, 2018).

D. Adequate Staffing:

- **CERT-RMM** and **C2M2** score highly (4/5) as they provide clear guidelines for resource management, ensuring that organisations have adequate and skilled personnel to manage incidents effectively. These models address staffing needs by integrating resource management with overall organisational resilience (Caralli et al., 2011; DOE, 2014).
- **ENISA IM3** scores well (4/5), particularly for CSIRTs, where staffing and resource allocation are critical to incident management. It emphasises the need for well-trained and adequately staffed teams (ENISA, 2022).

- **ISO/IEC 27035**, **NIST CSF**, and **CMMC** score slightly lower (3/5) as they address staffing more broadly, without the same level of detail or focus on maintaining staffing levels specific to incident management (Line, 2016; NIST, 2018; DoD, 2020).

E. **Information Security Governance:**

- **CERT-RMM**, **NIST CSF**, and **CMMC** score the highest (5/5) due to their comprehensive approaches to integrating information security governance within the overall cybersecurity strategy. These models provide robust frameworks for aligning security practices with organisational goals, ensuring strong governance structures (Caralli et al., 2011; NIST, 2018; DoD, 2020).
- **ISO/IEC 27035** and **ENISA IM3** score slightly lower (4/5). While they provide good governance practices, their primary focus remains on incident response rather than broader governance (Line, 2016; ENISA, 2022).
- **C2M2** also scores well (4/5), offering solid governance capabilities within its broader capability maturity framework, though it may require more organisational commitment to implement fully (DOE, 2014).

F. **Internal and External Communication:**

- **ENISA IM3** leads in this area (5/5) due to its strong emphasis on communication during incidents, particularly within CSIRTs. It ensures that internal and external stakeholders are informed and coordinated, which is crucial for effective incident management (ENISA, 2022).
- **CERT-RMM** and **ISO/IEC 27035** score well (4/5) as they provide clear guidelines for establishing communication protocols during incidents, ensuring that information flows effectively across the organisation (Caralli et al., 2011; Line, 2016).
- **NIST CSF**, **C2M2**, and **CMMC** score lower (3/5) as they address communication more broadly within the context of incident response without the same level of focus on detailed communication strategies specific to incidents (NIST, 2018; DOE, 2014; DoD, 2020).

G. **Information Security Culture:**

- CERT-RMM, NIST CSF, C2M2, CMMC, and ENISA IM3 all score well (4/5) as they each emphasise the importance of fostering a solid information security culture. These models integrate cultural development within their broader frameworks, promoting continuous improvement and a proactive security mindset across the organisation (Caralli et al., 2011; NIST, 2018; DOE, 2014; DoD, 2020; ENISA, 2022).
- ISO/IEC 27035 scores slightly lower (3/5) as it focuses more on incident handling than developing a broader security culture. However, it still provides some guidance on cultural aspects related to incident response (Line, 2016).

This comparison would give organisations the option to choose any combination of CMMs when conducting maturity assessments of specific areas or capabilities or opt to use only one CMM because of regulatory compliance objectives like:

- Organisations seeking ISO/IEC 27001 certification are often expected to follow the guidelines provided in ISO/IEC 27035 as part of their Information Security Management System (ISMS). Compliance with these standards is often a contractual or regulatory requirement in many industries, particularly those handling sensitive data or operating in highly regulated sectors.
- Organisations seeking to do business with the U.S. Department of Defense. CMMC is specifically designed as a compliance framework. It was developed by the U.S. Department of Defense (DoD) to ensure that Defence Industrial Base (DIB) contractors meet specific cybersecurity standards. CMMC is mandatory for companies seeking to engage in business with the U.S. Department of Defense.

Illustrated in the two tables below are examples of selecting MMCs. The first table shows the selection of MMCs with the highest scores in each area, while the next shows the selection of one MMC for complying with ISO/IEC 27001.

| Area of Capability / Model | Risk Management | Cyber Security Incident Handling Best Practices | Training and Awareness Around Incident Handling | Adequate Staffing | Information Security Governance | Internal and External Communication | Information Security Culture within the Organisation |
|---|---|---|---|---|---|---|---|
| **CERT-RMM** | ✓ | ✓ | ✗ | ✓ | ✓ | ✗ | ✓ |
| **ISO/IEC 27035** | ✗ | ✗ | ✗ | ✗ | ✗ | ✗ | ✗ |
| **NIST CSF** | ✗ | ✗ | ✗ | ✗ | ✗ | ✗ | ✗ |
| **C2M2** | ✗ | ✗ | ✗ | ✗ | ✗ | ✗ | ✗ |
| **CMMC** | ✗ | ✗ | ✗ | ✗ | ✗ | ✗ | ✗ |
| **IM3** | ✗ | ✗ | ✓ | ✗ | ✗ | ✓ | ✗ |

*Table 2: An example of selecting a combination of CMMs when assessing different capabilities based on the MMCs with the highest scores in each area*

| Area of Capability / Model | Risk Management | Cyber Security Incident Handling Best Practices | Training and Awareness Around Incident Handling | Adequate Staffing | Information Security Governance | Internal and External Communication | Information Security Culture within the Organisation |
|---|---|---|---|---|---|---|---|
| **CERT-RMM** | ✗ | ✗ | ✗ | ✗ | ✗ | ✗ | ✗ |
| **ISO/IEC 27035** | ✓ | ✓ | ✓ | ✓ | ✓ | ✓ | ✓ |
| **NIST CSF** | ✗ | ✗ | ✗ | ✗ | ✗ | ✗ | ✗ |
| **C2M2** | ✗ | ✗ | ✗ | ✗ | ✗ | ✗ | ✗ |
| **CMMC** | ✗ | ✗ | ✗ | ✗ | ✗ | ✗ | ✗ |
| **IM3** | ✗ | ✗ | ✗ | ✗ | ✗ | ✗ | ✗ |

*Table 3: An example of selecting one CMM when assessing different capabilities because of regulatory compliance purposes (ISO/IEC 27001)*

### 2.3 Theoretical Foundation for the Prioritisation of Incidents According to IRM Capability Levels

This sub-section addresses one of the core objectives of this study, which is manifested in a situation where an organisation conducts a maturity assessment of its capability level in any one of the socio-organisational areas, finds that it falls on the lower end of the threshold, and works on improving it. Meanwhile, an incident occurs during the improvement of the capability level of the area assessed. What priority should be assigned to the incident to minimise its impact?

Prioritising the incident response based on the outcome of IRM assessments involves a systematic approach to enhance incident prioritisation and response. This process would help organisations leverage their maturity assessment results to improve their capabilities and more effectively categorise, prioritise, and respond to incidents. The process can be illustrated in the following snapshot of a flow chart and broken down into five key steps listed next:

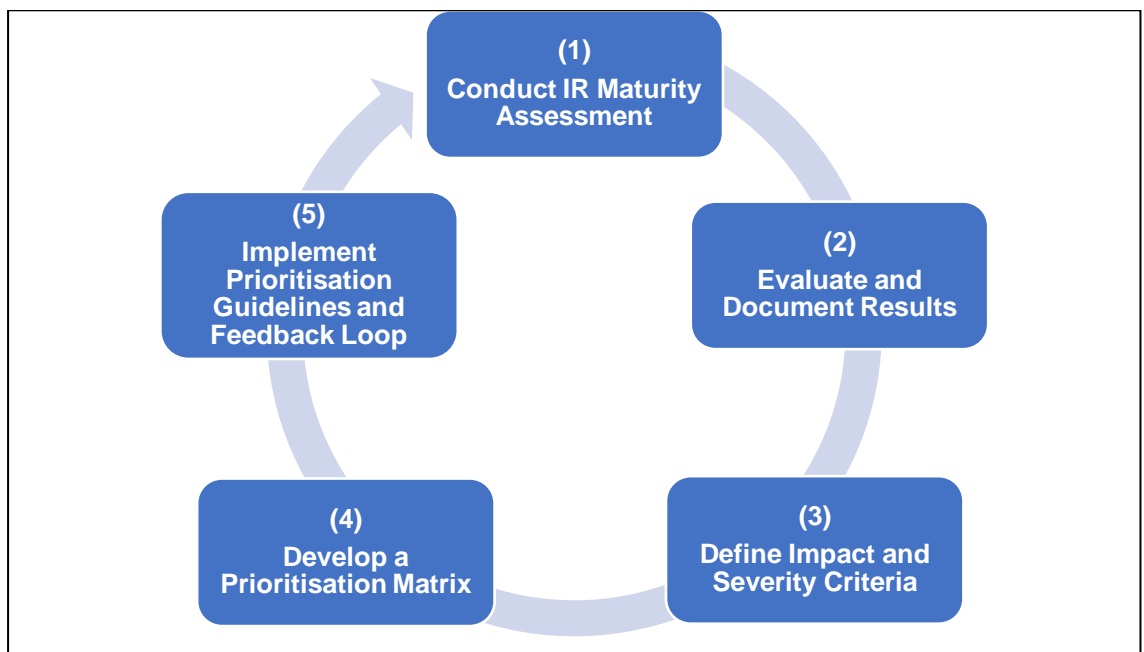

*Figure 1: The continuous process of integrating IRM and incident prioritisation*

Step 1 - Conducting a Maturity Assessment

Organisations would conduct a maturity assessment of the seven socio-organisational areas identified earler using any combination of the CMMs reviewed. This assessment evaluates the current state of the organisation's incident response capabilities, identifying strengths and weaknesses.

Below is a summary table of capability levels for the CMMs reviewed:

| CMM | Capability Level | Description |
|---|---|---|
| ENISA's IM3 | Ad-hoc | There is no formal incident management; responses are improvised. |
| | Defined | Basic processes are defined and documented; reactive response. |
| | Managed | Processes are managed and consistent; performance is measured. |
| | Controlled | Incident management is controlled and proactive; metrics are used. |
| | Optimised | Continuous improvement through proactive management and optimisation. |
| CMMC | Ad-hoc | Practices are performed in an inconsistent and sometimes ad-hoc manner. |
| | Reactive | Practices are performed reactively, often triggered by events. |
| | Defined | Practices are documented and standardised across the organisation. |
| | Managed | Practices are managed and measured for effectiveness and efficiency. |
| | Optimised | Continuous improvement processes are in place, and practices are optimised. |
| C2M2 | Ad-hoc | There are no formal incident response practices; responses are improvised. |
| | Initial | Initial, unstructured approach; processes are reactive. |
| | Repeatable | Basic processes are documented, and there is repeatability in response actions. |
| | Defined | Processes are well-defined, documented, and communicated. |
| | Managed | Processes are managed and monitored; performance is measured. |
| | Optimised | Continuous improvement through feedback and process optimisation. |
| CSF | Partial | Practices are not formalized; risk management is ad-hoc. |
| | Risk Informed | Practices are repeatable; risk management is risk informed. |

| CMM | Capability Level | Description |
|---|---|---|
| | Repeatable | Practices are consistent across the organisation and formally documented. |
| | Adaptive | Practices are managed, measurable, and adaptive to changing threats. |
| | Optimised | Continuous improvement and innovation in practices. |
| **CERT-RMM** | Ad-hoc | There are no formal practices; incident response is reactive and uncoordinated. |
| | Initial | Basic practices are in place; responses are still reactive. |
| | Managed | Practices are managed and consistent; performance is monitored. |
| | Defined | Practices are well-defined and documented; processes are proactive. |
| | Quantitatively Managed | Processes are quantitatively managed; metrics are used for decision-making. |
| | Optimised | Continuous improvement through innovative practices and feedback loops. |
| **ISO/IEC 27035** | Ad-hoc | Incident response is reactive; there are no formal processes. |
| | Initial | Initial processes are established but not fully documented. |
| | Repeatable | Processes are repeatable and documented; basic coordination. |
| | Managed | Processes are managed and monitored; incident response is coordinated. |
| | Optimised | Continuous improvement in processes through regular review and updates. |

*Table 4: Summary of Capability Levels for IR CMMs*

**Explanation of Capability Levels**

- Ad-hoc: Practices are informal, unstructured, and not standardised across the organisation. Responses are typically improvised, and there is little consistency in how incidents are managed.
- Reactive/Initial: Some basic processes are in place, often in response to specific events. However, these practices are not consistently documented or followed.
- Repeatable/Risk Informed: Processes are documented and repeatable. The organisation begins to standardise its practices and make them consistent across different departments.
- Defined: Practices are well-documented, standardised, and communicated throughout the organisation. Processes are proactive rather than reactive.
- Managed: Incident response practices are managed and monitored for effectiveness. Performance metrics are used to ensure that processes are followed correctly and identify improvement areas.
- Controlled/Quantitatively Managed: Incident management is controlled and proactive. Processes are quantitatively managed, and metrics are used for decision-making.
- Optimised: The organisation has a culture of continuous improvement, with feedback loops and innovation driving the optimisation of incident response practices.

Based on the explanation above, a relationship between the different maturity models is observed, and they can be aligned to a certain extent based on the similarities in their capability levels. The alignment allows organisations to translate their maturity levels across different models, providing a unified understanding of their incident response maturity.

| Capability Maturity Model | | | | | |
|---|---|---|---|---|---|
| C2M2 | CERT-RMM | ISO/IEC 27035 | ENISA's IM3 | CSF | CMMC |

| Capability Levels | | | | | | | |
|---|---|---|---|---|---|---|---|
| | **Ad-hoc** | Ad-hoc | Ad-hoc | Ad-hoc | Ad-hoc | Partial | Ad-hoc |
| | **Reactive** | Initial | Initial | Initial | Defined | Risk Informed | Reactive |
| | **Repeatable** | Repeatable | Managed | Repeatable | Managed | Repeatable | Defined |
| | **Proactive** | Defined | Defined | Managed | Controlled | Adaptive | Managed |
| | **Optimised** | Managed | Quantitatively Managed | Optimised | Optimised | Optimised | Optimised |
| | | Optimised | Optimised | | | | |

*Table 5: Alignment of Capability Levels Across Maturity Models*

**Explanation of the Alignment**

- Ad-hoc:
    - Practices are informal, unstructured, and not standardised. Responses are typically improvised.
    - Equivalent Levels: **Ad-hoc** (C2M2, CERT-RMM, ISO/IEC 27035, ENISA's IM3, CMMC), **Partial** (CSF).

- Reactive:
    - Basic processes are established but not fully documented or consistently followed. Responses are still reactive.
    - Equivalent Levels: **Initial** (C2M2, CERT-RMM, ISO/IEC 27035), **Defined** (ENISA's IM3), **Risk Informed** (CSF), **Reactive** (CMMC).

- Repeatable:
    - Processes are documented and repeatable, standardised across the organisation, and followed consistently.
    - Equivalent Levels: **Repeatable** (C2M2, ISO/IEC 27035, CSF), **Managed** (CERT-RMM, ENISA's IM3), **Defined** (CMMC).

- Proactive:
    - Practices are well-documented, proactive, managed, and monitored for effectiveness.
    - Equivalent Levels: **Defined** (C2M2, CERT-RMM), **Controlled** (ENISA's IM3), **Adaptive** (CSF), **Managed** (ISO/IEC 27035, CMMC).

- Optimised:
    - Incident management is controlled and proactive, with continuous improvement through feedback and optimisation.
    - Equivalent Levels: **Managed/Optimised** (C2M2), **Quantitatively Managed/Optimised** (CERT-RMM), **Optimised** (ISO/IEC 27035, ENISA's IM3, CSF, CMMC).

Step 2 - Evaluate and Document Results

This step aims to document and analyse the results of the maturity assessments conducted in Step 1, clearly understanding the organisation's current incident response capabilities across various areas of interest.

For each capability area assessed, the maturity level determined in the assessment should be documented, including any relevant qualitative information or observations that provide context.

To better understand the organisation's current state of readiness and capability in each area, the following activities are carried out:

- Compare Against Benchmarks: The documented maturity levels should be compared against industry benchmarks or best practices to understand how the organisation stands relative to peers.
- Identify Strengths and Weaknesses: Highlighting areas where the organisation is performing well (high maturity levels) and areas that need improvement (low maturity levels).
- Qualitative Analysis: Any qualitative observations or notes that provide additional insights into the quantitative maturity levels should also be considered. For example, a capability area rated "Reactive" might have specific challenges to address.

Based on the alignment of CMMs' capability levels suggested in Step 1, the scoring scale below would indicate the capability maturity level given through any CMM.

|  |  | CMMs | | | | | | Score |
|--|--|------|--|--|--|--|--|-------|
|  |  | C2M2 | CERT-RMM | ISO/IEC 27035 | ENISA's IM3 | CSF | CMMC | |
| Capability Levels | Ad-hoc | Ad-hoc | Ad-hoc | Ad-hoc | Ad-hoc | Partial | Ad-hoc | 1 |
| | Reactive | Initial | Initial | Initial | Defined | Risk Informed | Reactive | 2 |
| | Repeatable | Repeatable | Managed | Repeatable | Managed | Repeatable | Defined | 3 |
| | Proactive | Defined | Defined | Managed | Controlled | Adaptive | Managed | 4 |
| | Optimised | Managed | Quantitatively Managed | Optimised | Optimised | Optimised | Optimised | 5 |
| | | Optimised | Optimised | | | | | |

*Table 6: Suggested scoring for aligned capability levels across CMMs*

The following is an example of the outcome of a maturity assessment using a combination of CMMs:

| CMM Area | CMM Used | Maturity Level | Score |
|----------|----------|----------------|-------|
| Risk Management | **CMMC** | Ad-hoc | 2 |
| Incident Handling Best Practices | **CMMC** | Reactive | 2 |
| Training and Awareness | **CERT-RMM** | Repeatable | 3 |
| Adequate Staffing | **ISO/IEC 27035** | Reactive | 2 |
| InfoSec Governance | **CERT-RMM** | Optimised | 5 |
| Communication | **ENISA's IM3** | Ad-hoc | 1 |

| CMM Area | CMM Used | Maturity Level | Score |
|---|---|---|---|
| InfoSec Culture | **NIST CSF** | Proactive | 4 |

*Table 7: An example of scoring the outcome of an assessment using a combination of CMMs*

The following is an example of the outcome of a maturity assessment using ISO/IEC 27035:

| CMM Area | CMM Used | Maturity Level | Score |
|---|---|---|---|
| Risk Management | ISO/IEC 27035 | Ad-hoc | 1 |
| Incident Handling Best Practices | ISO/IEC 27035 | Reactive | 2 |
| Training and Awareness | ISO/IEC 27035 | Reactive | 2 |
| Adequate Staffing | ISO/IEC 27035 | Reactive | 2 |
| InfoSec Governance | ISO/IEC 27035 | Proactive | 4 |
| Communication | ISO/IEC 27035 | Reactive | 2 |
| InfoSec Culture | ISO/IEC 27035 | Optimised | 5 |

*Table 8: An example of scoring the outcome of an assessment using the same CMM*

Calculating the average score of the capabilities can serve as a baseline to compare against future capability levels of future IR capability maturity assessments and realise improvements or declines in the levels of capabilities.

If capability maturity scores are taken from the previous table, for example, the average capability score of the organisation would look like the following:

Average Capability Score = $\frac{1+2+2+2+4+2+5}{7} \approx 2.57$

Identifying specific areas where improvements are needed to enhance the organisation's incident response capabilities should also be part of the evaluation of the assessment results, which includes the following activities:

- <u>Gap Analysis:</u> Perform a gap analysis to determine the difference between current and desired maturity levels in each area.
- <u>Prioritise Improvements:</u> Based on the gap analysis, prioritise areas for improvement. Focus on areas with the lowest maturity levels or critical for incident response.
- <u>Action Plan Development:</u> Develop an action plan to address identified gaps. This plan should include specific actions, responsible parties, timelines, and resources needed.

### Step 3 - Define Impact and Severity Levels

Although now that the maturity levels of each capability are determined, they cannot be applied to every incident type. The impact and severity of each incident type are different. For example, the impact of phishing attacks might be rated lower in immediate operational disruption compared to a DDoS attack but higher in terms of severity and long-term impact, as they can lead to a range of follow-on attacks and possibly data breaches.

Impact and severity levels need to be clearly defined to ensure consistent and accurate prioritisation of incidents across different incident types. To achieve that, clear and consistent criteria must be established for evaluating the impact and severity of incidents.

Impact criteria would define what constitutes low, medium, and high impact. Factors such as operational disruption, financial loss, data sensitivity, reputational damage, and legal/regulatory impact should be considered here.

Severity criteria, on the other hand, define what constitutes low, moderate, high, and critical severity. Factors such as the scope of the incident, ease of detection and containment, potential for escalation, duration and recovery time, and impact on customers and partners are considered.

**Impact Levels Criteria and Scores:**

| Impact Level | Criteria | Score |
|---|---|---|
| Low | Minimal disruption, negligible financial impact, and involves non-sensitive data. | 1 |
| Medium | Partial disruption, moderate financial impact, and involves sensitive data. | 2 |
| High | Significant disruption, substantial financial loss, and involves highly sensitive data. | 3 |

*Table 9: Impact levels, their criteria, and scores*

**Severity Levels Criteria and Scores:**

| Impact Level | Criteria | Score |
|---|---|---|
| Low | Affects a small number of systems/users and is easily containable. | 1 |
| Moderate | Affects multiple systems/departments and is manageable with existing resources. | 2 |
| High | Affects significant portions of infrastructure and requires significant resources. | 3 |
| Critical | Affects the entire organisation and is highly challenging to contain and recover. | 4 |

*Table 10: Severity levels, their criteria, and scores*

### Step 4 - Develop a Prioritisation Matrix

A prioritisation matrix could help systematically prioritise incidents between iterations of IR capability maturity assessments. The matrix would rely on the three elements gather thus far: namely, the impact and severity scores, and the average capability score.

Developing a priority matrix ensures that the most critical incidents receive immediate attention and resources. The underlying idea aligns with established principles in risk management and cybersecurity. These principles include the balancing of risks (impact and severity) with mitigation efforts (capability).

**Priority Score** = $\frac{(Impact\ Score + Severity\ Score)}{Average\ Capability\ Score}$

In the formula above, the sum of the impact and severity scores is divided by the average capability score. If the average capability is high, it will lower the overall "Priority Score," indicating that the high impact and severity are mitigated by strong "mature" capabilities. Conversely, if the average capability is low, the "Priority Score" will remain high, indicating that the organisation is less able to mitigate the incident's impact and severity. Here is an example of a calculation for a phishing attack type of incident.

The maximum Priority Score occurs when Impact + Severity = 7 and Capability = 1, which would be 7, while the minimum Priority Score occurs when Impact + Severity = 2 (assuming both impact and severity scores are at least 1) and Capability = 5, which would give a Priority Score of 0.4.

Given that the Priority Score ranges from 0.4 to 7, the following priority levels can be derived:

| Threshold | Criteria | Level |
|---|---|---|
| 0.4-2.0 | Scores in this range represent situations where high capability effectively mitigates the impact and severity. | Low |
| 2.1-3.5 | Scores here indicate a moderate level of risk where capability is somewhat mitigating the impact, but not entirely. | Medium |
| 3.6-5.0 | Scores suggest that the impact and severity are significant, and capability is not enough to reduce the risk substantially. | High |
| 5.1-7.0 | Scores occur when high impact and severity are not adequately mitigated by capability, representing the highest level of risk. | 4 |

*Table 11: Thresholds corresponding to priority scores*

An example of calculating the priority score and level for a phishing attack:

- <u>Incident Type:</u> Phishing Attack
- <u>Impact:</u> Medium (2) → See details on how to derive this in Step 3
- <u>Severity:</u> High (3) → See details on how to derive this in Step 3
- <u>Overall Capability Score:</u> 2.57 → See details on how to derive this in Step 2

Priority Score = (Impact + Severity) / Capability
Priority Score = (2 + 3) / 2.57
Priority Score = 5 / 2.57
Priority Score ≈ 1.95

Similarly, if other incident types are included in the matrix to serve as a reference in case they occur, a prioritisation matrix for an organisation would look like the following:

| Incident Type | Impact | Severity | Capability | Priority Score | Priority Level |
|---|---|---|---|---|---|
| Phishing Attacks | 2 | 3 | 2.57 | 1.95 | Low |
| Zero-Day Exploits | 3 | 4 | 2.57 | 2.72 | Medium |
| Data Corruption | 3 | 2 | 2.57 | 1.95 | Low |
| DDoS Attacks | 1 | 2 | 2.57 | 1.17 | Low |
| Insider Threats | 2 | 3 | 2.57 | 1.95 | Low |
| Malware/Ransomware | 3 | 4 | 2.57 | 2.72 | Medium |
| Unauthorised Access | 3 | 3 | 2.57 | 2.33 | Medium |
| System Misconfigurations | 2 | 1 | 2.57 | 1.714 | Low |

*Table 12: An example of a prioritisation matrix*

It is crucial to ensure the accuracy and reliability of the prioritisation matrix through review and validation. Some actions that support this would be:

- <u>Peer Reviews:</u> Have the matrix reviewed by peers or other incident response team members to ensure consistency and accuracy.
- <u>Management Reviews:</u> Present the matrix to management for validation, approval, and continuous support.
- <u>Adjusting as Necessary:</u> Based on feedback from the review process, adjust the matrix to ensure it accurately reflects the organisation's priorities and capabilities.

Below is a more specific example on prioritising a ransomware attack targeting a multi-national organisation that has completed steps 1-3 for all of its four subsidiaries or branches globally, the prioritisation matrix would resemble the table below. Giving the organisation the ability to not only recognise that they need to provide more support to their branch in Singapore, but also an assurance of the resilience across all branches and subsidiaries against the attack.

| Branch/Subsidiary | Incident Type | Impact | Severity | Capability | Priority Score | Priority Level |
|---|---|---|---|---|---|---|
| London | Ransomware Attack | 3 | 4 | 2.17 | 3.23 | Medium |
| Paris | | | | 3.42 | 2.05 | Medium |
| Singapore | | | | 1.87 | 3.74 | High |
| Melbourne | | | | 3.02 | 2.32 | Medium |

*Table 13: An example of a prioritisation matrix for a multi-national organisation against a ransomware attack*

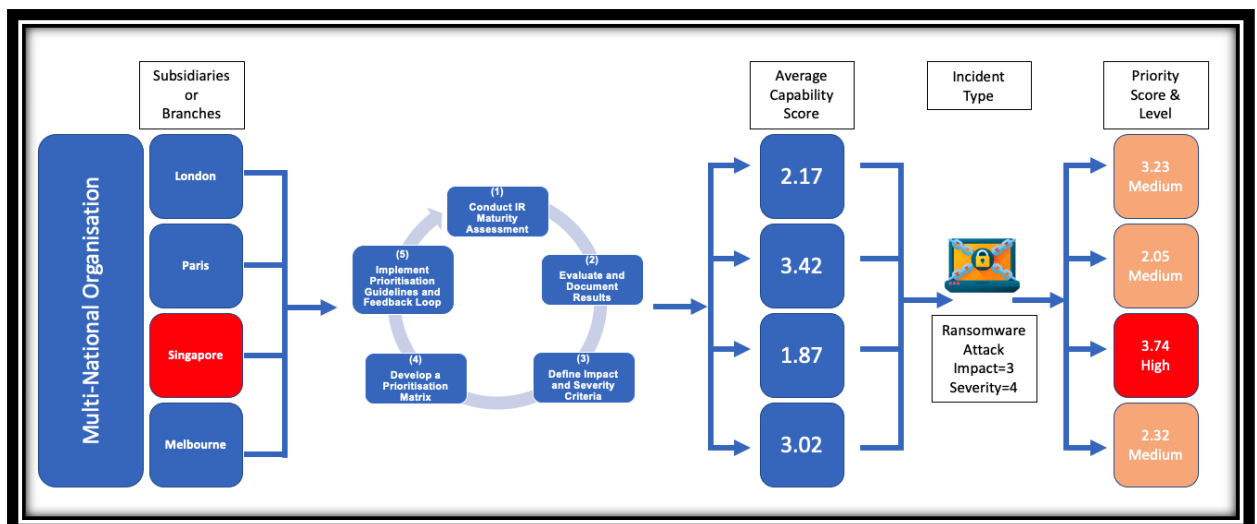

*Figure 2: An example of a prioritisation matrix for a multi-national organisation against a ransomware attack (attribution: ransomware icon from flaticon.com)*

### Step 5 - Implement Prioritisation Guidelines and Feedback Loop

Clear guidelines for applying the prioritisation process across the organisation should be established and enforced, starting with creating detailed guidelines that outline how the prioritisation matrix should be used to classify and respond to incidents, which includes specific instructions on how to interpret priority scores and assign priority levels. These guidelines should be maintained up-to-date and made easily accessible to all stakeholders; they should also be periodically reviewed to reflect changes in the organisation's environment, threat landscape, or incident response capabilities.

Enforcing the guidelines could be achieved through communication and training. All relevant personnel (including incident response teams and management) should be made aware of and understand the prioritisation guidelines. Training sessions could be provided to incident response teams on applying the prioritisation guidelines effectively, which will aid in enforcement.

Using IR capability maturity assessments to prioritise incidents should not be a one-time activity. A feedback loop should be established to continuously update the prioritisation matrix and guidelines based on new data and evolving threats. Regular reviews, updates, and IR capability maturity assessments would reflect changes in the organisation's capabilities and the threat landscape. This continuous improvement ensures that the incident prioritisation remains effective and aligned with the organisation's current state.

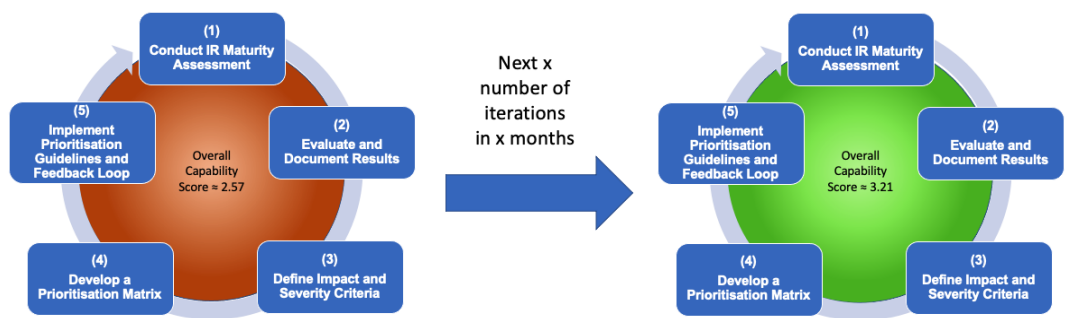

*Figure 3: Regular IR capability maturity assessments could improve the overall IR capability of an organisation.*

3. Discussion

Integrating maturity assessments with incident prioritisation allows for a more systematic and effective response to incidents. Using CMMs to assess those factors continually to improve them is crucial while using the outcomes of assessments to prioritise responses to incidents that happen between iterations of assessments to ensure that even if any of the factors is not optimised, the organisation would still be able to recognise the impact and act more efficiently on minimising the impact.

From a policy perspective, organisations need to develop and enforce robust incident response policies that are regularly reviewed and updated. These policies should define the roles and responsibilities of all stakeholders, outline the procedures for incident detection, response, and recovery, and include guidelines for regular testing and improvement. Adequate resource allocation is also crucial; as organisations invest in advanced detection and response technologies, they should retrospectively ensure that their incident response teams are adequately staffed, trained, and supported to use them properly.

For researchers, this study highlights the need for further exploration into the specific human and organisational factors that influence incident response effectiveness. Future research should consider comparative analyses of different maturity models and their effectiveness in various organisational contexts, which would provide valuable insights.

4. Conclusion and Future Work

The significance of this research lies in its thorough examination of the intersection between IRM assessments and incident prioritisation. The study addresses a crucial aspect often overlooked in technical cybersecurity solutions by focusing on human and organisational factors. The findings emphasise that technological advancements alone cannot mitigate cybersecurity risks; human and organisational readiness is equally vital.

This research contributes to the existing body of knowledge by providing a detailed analysis of how IR CMMs can be applied by focusing on socio-organisational aspects to improve incident prioritisation.

Future research could also focus on longitudinal studies to assess the long-term impact of incident response improvements on organisational resilience (Zare et al., 2022). Comparative analyses of different maturity models and their effectiveness in various organisational contexts would provide valuable insights for practitioners (Rao & Nayak, 2017). Further research is needed to explore the specific human and organisational factors influencing incident response effectiveness, particularly in different industry sectors and organisational sizes (Zare et al., 2022).

The following are recommendations for further research based on this project's findings, including specific research questions, methodologies, and areas of investigation.

- Developing Integrated Frameworks: Future research could focus on developing frameworks explicitly linking IRM assessments with prioritisation processes. This research could involve creating standardised templates or tools that guide organisations on the most efficient use of maturity assessment results to prioritise incidents effectively.

- Algorithmic Approaches: Creating algorithms that translate maturity model scores into incident prioritisation metrics could provide a more automated and objective approach to incident management. These algorithms could consider various factors such as risk, impact, and organisational priorities.

- Validation Through Case Studies: Detailed case studies of organisations implementing integrated approaches can provide valuable insights. These case studies could explore the challenges and benefits of linking maturity assessments with incident prioritisation and offer best practices.

- Guidelines and Best Practices: Developing comprehensive guidelines outlining best practices for integrating IRM assessments with prioritisation strategies could help bridge the gap. These guidelines should be practical and based on empirical evidence.

- Adaptive Incident Management Models: Research could explore the development of adaptive incident management models evolving with an organisation's maturity level. These models would ensure that incident prioritisation processes align with current capabilities and threat landscape.

These recommendations emphasise the need for ongoing evaluation and improvement of incident response strategies, which aligns with the dynamic nature of cybersecurity threats and organisational responses.

References

Bajak, A. (2018). Equifax: Inside the Hack and the Fallout. Retrieved from https://www.triaxiomsecurity.com/key-lessons-learned-from-the-equifax-data-breach

Baskerville, R., Spagnoletti, P., & Kim, J. (2014). Incident-Centered Information Security: Managing a Strategic Balance between Prevention and Response. https://www.researchgate.net/publication/259129309_Incident-Centered_Information_Security_Managing_a_Strategic_Balance_between_Prevention_and_Response

Bitzer, M., Häckel, B., Leuthe, D., Ott, J., Stahl, B., & Strobel, J. (2023). Managing the inevitable–A maturity model to establish incident response management capabilities. *Computers & Security, 125*, 103050. https://www.sciencedirect.com/science/article/abs/pii/S0167404823000502

Caralli, R. A., White, D. W., & Allen, J. H. (2011). The CERT resilience management model : a maturity model for managing operational resilience (1st edition). Addison Wesley. Retrieved from https://learning.oreilly.com/library/view/cert-r-resilience-management/9780132565905/

Chen, A., Pavlou, P. A., & Zhang, X. (2018). The Equifax breach: A case study in escalating negative externality. *MIS Quarterly.* https://misq.org

Cimpanu, C. (2018). Equifax hack blamed on known web flaw, oversight and a passive IT department. Retrieved from https://www.zdnet.com/article/equifax-hack-blamed-on-known-web-flaw-oversight-and-a-passive-it-department/

DoD. (2020). Cybersecurity maturity model certification (CMMC). U.S. Department of Defense. Retrieved from https://www.acq.osd.mil/cmmc/

DOE. (2014) Cybersecurity capability maturity model (C2M2). U.S. Department of Energy. Retrieved from https://www.energy.gov/ceser/activities/cybersecurity-critical-energy-infrastructure/cybersecurity-capability-maturity

DeSimone, A. & Horton, N. (2017). SONY'S NIGHTMARE BEFORE CHRISTMAS The 2014 North Korean Cyber Attack on Sony and Lessons for US Government Actions in Cyberspace. Johns Hopkins Applied Physics Laboratory. Retrieved from https://apps.dtic.mil/sti/pdfs/AD1046744.pdf

Elsevier. (n.d.). ScienceDirect. Retrieved from https://www.sciencedirect.com/

ENISA. (2022). ENISA CSIRT maturity framework - Updated and improved. European Union Agency for Cybersecurity. Retrieved from https://www.enisa.europa.eu/publications/enisa-csirt-maturity-framework

Gallagher, S. (2014). Sony Pictures hacking: What we know and what could happen next. Retrieved from https://www.washingtonpost.com/news/the-switch/wp/2014/12/04/sony-pictures-hacking-what-we-know-and-what-could-happen-next/

GAO, U.S. Government Accountability Office (2018). Data Protection: Actions Taken by Equifax and Federal Agencies in Response to the 2017 Breach. Retrieved from https://www.gao.gov/assets/700/694347.pdf

Google. (n.d.). Google Scholar. Retrieved from https://scholar.google.com/
21

Gordon, L. A., Loeb, M. P., Lucyshyn, W., & Richardson, R. (2017). CSI/FBI computer crime and security survey. *Computer Security Institute.* https://www.researchgate.net/publication/243784811_CSIFBI_Computer_Crime_and_Security_Survey

Hartnett, T. (2017). The Equifax Breach: Analysis and Recommendations. Retrieved from https://sevenpillarsinstitute.org/case-study-equifax-data-breach/

IEEE. (n.d.). IEEE Xplore Digital Library. Retrieved from https://ieeexplore.ieee.org/

Line, M. B. (2015). UNDERSTANDING INFORMATION SECURITY INCIDENT MANAGEMENT PRACTICES, A case study in the electric power industry. Norwegian University of Science and Technology. Retrieved from https://ntnuopen.ntnu.no/ntnu-xmlui/bitstream/handle/11250/2359707/Line,%20Maria%20Bartnes.pdf?sequence=4

NIST. (2018). Framework for improving critical infrastructure cybersecurity. National Institute of Standards and Technology. Retrieved from https://www.nist.gov/cyberframework

Pappenheim da Silva, B., da Silva, A.J., Davidsen, J.E. (2019). Information Security Governance, Technology, Processes and People: Compliance and Organisational Readiness. In: Jahankhani, H., Kendzierskyj, S., Jamal, A., Epiphaniou, G., Al-Khateeb, H. (eds) Blockchain and Clinical Trial. Advanced Sciences and Technologies for Security Applications. Springer, Cham. https://doi.org/10.1007/978-3-030-11289-9_4

Perlroth, N. (2014). Hackers' attack cracked 7% of passwords at Sony Pictures. *The New York Times.* https://www.nytimes.com/2014/12/04/technology/sony-attack-exposed-passwords.html

Rao, S., & Nayak, R. (2017). IRM models: A comparative study. *US Journal of Information Security and Applications, 34*, 1-10. https://www.sciencedirect.com/science/article/abs/pii/S2214212616301287

Rieger, D., Tjoa, S. (2018). A Readiness Model for Measuring the Maturity of Cyber Security Incident Management. In: Xhafa, F., Barolli, L., Greguš, M. (eds) Advances in Intelligent Networking and Collaborative Systems. INCoS 2018. Lecture Notes on Data Engineering and Communications Technologies, vol 23. Springer, Cham. https://doi.org/10.1007/978-3-319-98557-2_26

Saleem, H. & Naveed, M. (2020). SoK: Anatomy of Data Breaches, Proceedings on Privacy Enhancing Technologies. *Sciendo*. Retrieved from https://www.researchgate.net/publication/346937548_SoK_Anatomy_of_Data_Breaches#full-text

Schein, E. H. & Schein, P.A. (2016). *Organisational culture and leadership, 5th Edition.* John Wiley & Sons. https://www.wiley.com/en-us/Organisational+Culture+and+Leadership%2C+5th+Edition-p-9781119212041

Shu, X., Tian, K., Ciambrone, A., & Yao, D. (2017). Breaking the Target: An analysis of Target data breach and lessons learned. Retrieved from https://arxiv.org/abs/1701.04940

Siponen, M., & Vance, A. (2010). Neutralisation: New insights into the problem of employee information systems security policy violations. *MIS Quarterly.* https://www.researchgate.net/publication/279550478_Neutralization_New_Insights_into_the_Problem_of_Employee_Information_Systems_Security_Policy_Violations#full-text
22